\documentclass[journal]{IEEEtran}
\usepackage{amsmath,amsfonts,amssymb,bm} 
\usepackage{array,stfloats,graphicx}
\usepackage{url,cite,siunitx}
\usepackage[hidelinks,colorlinks=true,allcolors=blue]{hyperref}
\def\BibTeX{{\rm B\kern-.05em{\sc i\kern-.025em b}\kern-.08em
    T\kern-.1667em\lower.7ex\hbox{E}\kern-.125emX}}
\usepackage{balance}

\begin{document}

\title{Effect of interfaces on supercurrent through ferromagnetic materials}
\author{Swapna Sindhu Mishra, Reza Loloee, and Norman O. Birge\\
Department of Physics and Astronomy, Michigan State University, East Lansing, MI 48824, USA
\thanks{Manuscript received October 19, 2022; revised June 14, 2023. 
Corresponding author: Norman Birge (email: birge@msu.edu).}}
\date{\today}
\maketitle

\begin{abstract}
Ferromagnetic Josephson junctions exhibit fascinating physics and the potential for applications in superconducting logic and memory. The junctions in a demonstrated superconducting memory prototype contain a magnetic spin-valve structure with Ni as the fixed layer and NiFe (Permalloy) as the free layer. However, NiFe exhibits poor supercurrent transmission, which limits the efficiency of the Josephson junction. We have previously shown that the supercurrent transmission through a Cu/NiFe/Cu trilayer can be improved by adding thin layers of Ni between the Cu and NiFe -- possibly due to the advantageous spin-dependent transport properties of the Cu/Ni interfaces. In this work we explore this idea further by replacing the Cu/NiFe interfaces with Pd/NiFe, which also have more desirable transport properties. Compared to the reference junctions containing Cu/NiFe interfaces, the new junctions exhibit an increase in the $\pi$-state supercurrent by a factor of 2 along with a change in the position of the first $0-\pi$ transition.
\end{abstract}

\section{Introduction}
Ferromagnetic Josephson junctions are the topic of continued interest over the last few decades because of the interesting physics that arises from the interplay between competing forms of order at the ferromagnet(F)-superconductor(S) interface\cite{Buzdin2005}. They also have potential applications in superconducting computing (digital logic and memory)\cite{Ryazanov2012,Soloviev2017} and in various qubit designs\cite{Ioffe1999, Blatter2001, Yamashita2005, Feofanov2010}. The reasons for pursuing a superconducting computer are multi-fold: they can be more energy efficient than Si-based computers even after the cooling costs are included\cite{Holmes2013} and they can also be potentially used as a compact controller for a quantum computer\cite{Howington2019, Howe2022}. The interesting physics in these devices arises from the exchange splitting between the majority and minority spin bands in F layers. This causes the spin-singlet Cooper pairs to undergo rapid phase oscillations and decay in the F layer\cite{Demler1997,Buzdin2005}. In a ferromagnetic Josephson junction, the ground-state phase difference across the junction can be either 0 or $\pi$ depending on the thickness of the F layer due to these phase oscillations\cite{Ryazanov2001,Kontos2002}.

While $\pi$-junctions are interesting in their own  right, new opportunities arise if the ground-state junction phase can be toggled between the 0 and $\pi$ states.  This can be done by replacing the ferromagnetic layer with two different ferromagnets in a pseudo spin-valve configuration: one with a fixed magnetization direction, and the other whose magnetization direction can be toggled between parallel and antiparallel to the first\cite{Bell2004,Baek2014,Qader2014}. By carefully choosing the thicknesses of the two layers, one of the magnetic states produces a 0-junction while the other produces a $\pi$-junction\cite{Golubov2002,Gingrich2016}. Such junctions are at the heart of Northrop Grumman's ``Josephson Magnetic Random Access Memory" (JMRAM) architecture, which was demonstrated using Ni as the fixed magnetic layer, NiFe as the free layer and Cu spacer layers located both between the two F layers and between the F and S layers\cite{Dayton2018}. While Ni has been shown to support excellent supercurrent transmission\cite{Baek2018}, the supercurrent through NiFe is smaller and decays rapidly with NiFe thickness\cite{Robinson2006,Glick2017}.

For many applications\cite{Dayton2018,Feofanov2010}, the ferromagnetic S/F/S Josephson junctions act as passive phase shifters, and always remain in the superconducting state. They are surrounded by conventional S/I/S Josephson junctions (where I is an insulator) which undergo switching during logic or memory read operations. In such circuits, the S/F/S junctions must have larger critical current ($I_c$) than the S/I/S junctions to avoid getting switched into the voltage state. It is possible to increase $I_c$ simply by increasing the lateral area of the junctions, but this is not desirable because the larger magnetic bits are likely to have multi-domain magnetic states which make the magnetic switching properties less reproducible. The goal of this study is to increase the critical current density, $J_c$, of Josephson junctions containing a NiFe layer. Our approach is to take information gleaned from Giant Magnetoresistance (GMR) studies of transport through interfaces in metallic multilayers. While the supercurrent through a Josephson junction is an equilibrium property, the critical supercurrent $I_c$ is influenced by the same bulk and interface scattering properties that determine the normal-state transport\cite{Ambegaokar1963,Kulik1975,Beenakker1991, Beenakker1992}. For example, it is known that NiFe has a very short mean free path for minority electrons and the NiFe/Cu interface has both a high specific resistance and a large spin-dependent resistance asymmetry\cite{Vila2000}. In a previous work, we demonstrated that adding thin layers of Ni between Cu/NiFe interfaces increased the critical currents in our junctions by up to 4 times\cite{Mishra2022}. We speculated that this is because the interfacial properties of Cu/Ni (when compared to Cu/NiFe) are favorable to Cooper pair transmission: lower values of interface resistance and spin-scattering asymmetry. However, theoretical studies are required to confirm the exact mechanisms behind this increase.

The spin-dependent transport properties of many ferromagnet-normal metal (F/N) interfaces have been measured by the Bass and Pratt group at Michigan State University and tabulated in review articles\cite{Pratt2009,Bass2016}. In this work, we replace each Cu spacer layer in our Josephson junctions with Pd. We chose this replacement because the properties of the Pd/NiFe interface appear more suitable for Cooper pair transmission than the Cu/NiFe interface, although not as good as the Cu/Ni interface discussed in Ref. \cite{Mishra2022}. Assuming the reason behind our motivation is correct, we expect this replacement to improve the critical current somewhat, although perhaps not as much as the addition of thin Ni between Cu/NiFe interfaces. The results of this study are promising: we find that we can increase the supercurrent transmission through Josephson junctions containing NiFe by about a factor of 2 in the $\pi$ state. The magnetic switching behavior of Pd/NiFe/Pd films is somewhat degraded compared to that of Cu/NiFe/Cu films, possibly because of the polarization of Pd, however the switching fields and coercivities are in a comparable range which should not impede in their use in applications such as cryogenic memory.

\section{Fabrication and Measurement}
\subsection{Thin films}
Thin films with a multilayer structure of Pd(2)/NiFe($d_{\mathrm{NiFe}}$)/Pd(2) (layer thicknesses are in nanometers) were deposited on an oxidized Si substrate with dc magnetron sputtering. The composition of our NiFe sputtering target is Ni$_{81}$Fe$_{19}$, but energy dispersive X-ray (EDX) analysis of sputtered thick films suggests a slightly different composition of Ni$_{82}$Fe$_{18}$. The deposition process was performed at an Ar pressure of \SI{0.3}{\Pa} and a substrate temperature around \SI{250}{\kelvin}. The base pressure of the sputtering chamber before the deposition was \SI{4e-6}{\Pa}. The NiFe layer thickness, $d_{\mathrm{NiFe}}$ was varied from \SI{0.4} to \SI{3.2}{\nm} in steps of \SI{0.4}{\nm}. Previously sputtered thin Cu(2)/NiFe($d_{\mathrm{NiFe}}$)/Cu(2) samples with the same $d_{\mathrm{NiFe}}$ variation were used for comparison. All of the above film layers were grown on top of a [Nb(25)/Al(2.4)]$_3$/Nb(20) base layer to match the base structure of our Josephson junctions, and were capped with Nb(5) to prevent oxidation. All samples were sputtered in the presence of a small magnetic field to orient the magnetic easy axis of the NiFe in the desired direction. The sputtering rates were: \SI{0.49}{nm/s} (Nb), \SI{0.19}{nm/s} (Al), \SI{0.27}{nm/s} (Pd), \SI{0.21}{nm/s} (NiFe), and \SI{0.36}{nm/s} (Au). The deposition process is highly reproducible from sample to sample and across different runs because of good stability in the sputter rates and computer control of the deposition time. Low-angle X-ray reflection measurements on films of various materials and polarized neutron reflectometry studies of superlattices confirm that the actual thicknesses are very close to the nominal deposition thicknesses obtained using the \textit{in-situ} crystal thickness monitor \cite{Quarterman2020}.

The moment versus field measurements for the above samples were performed at a temperature of \SI{10}{\kelvin} using a SQUID-based Vibrating Sample Magnetometer (VSM).

\subsection{Josephson junctions}
The fabrication process for our ferromagnetic Josephson junctions has been discussed in detail previously\cite{Glick2017}. First, the photo-lithographic stencil for the bottom lead was patterned on a clean Si substrate, and then [Nb(25)/Al(2.4)]$_3$/Nb(20)/Pd(2)/NiFe($d_{\mathrm{NiFe}}$)/Pd(2)/Nb(5) /Au(10) was sputtered where $d_{\mathrm{NiFe}}$ was varied from \SI{0.4}{} to \SI{3.2}{\nm} in steps of \SI{0.2}{\nm}. The Josephson junctions were then patterned by e-beam lithography with a negative ma-N2401 e-beam resist followed by ion-milling. These junctions are elliptical with lateral dimensions of \SI{1.25}{\micro \meter} $\times$ \SI{0.5}{\micro \meter} and their major axis is oriented along the magnetic easy axis set using a small magnetic field in the previous deposition step. After ion milling, the area surrounding the junctions was covered with $\mathrm{SiO_x}$ \textit{in-situ} to avoid electrical shorts between the bottom and top superconducting electrodes to be deposited next. The e-beam resist was then removed, and the top lead stencil was patterned using photolithography. \SI{5}{\nm} of the Au(10) capping layer from the previous deposition was ion milled \textit{in-situ} to improve surface contact and then the top Nb(150)/Au(10) superconducting electrodes were deposited by sputtering. 

Josephson junctions were mounted on a home-built probe and then inserted inside a liquid helium Dewar for transport measurements at \SI{4}{\kelvin}. This probe has a built-in superconducting magnet and is also equipped with a SQUID-based sensing device to measure very small critical currents with low voltage noise. $I-V$ curves for all junctions were measured in magnetic fields applied along the major axes of the elliptical junctions, up to fields of \SI{0.1}{\tesla} in both the positive and negative directions.

\section{Results}
\subsection{Thin film magnetics}
\begin{figure}[!htbp]
\includegraphics[width=\linewidth]{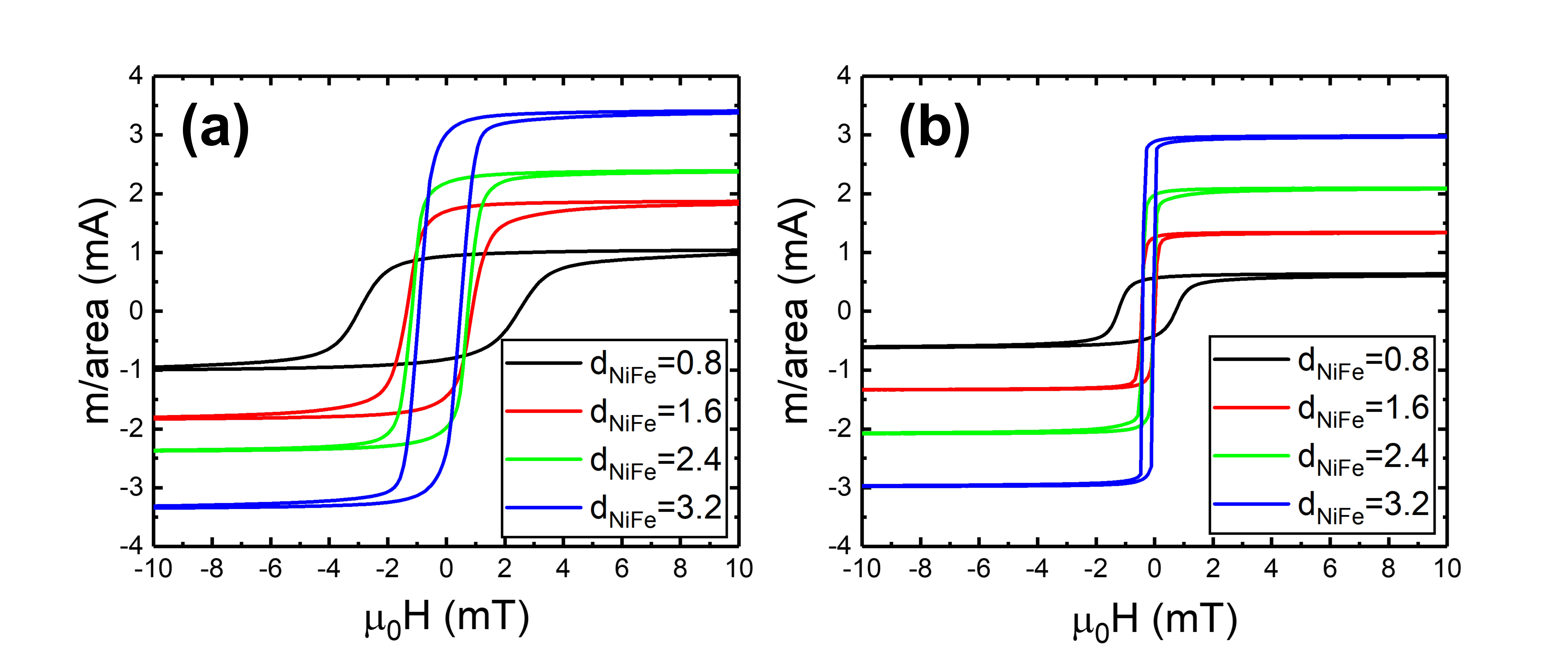}
\caption{$m/\mathrm{area}$ vs $H$ for selected (a) Pd(2)/NiFe($d_{\mathrm{NiFe}}$)/Pd(2) and (b) Cu(2)/NiFe($d_{\mathrm{NiFe}}$)/Cu(2) films measured at $T = \SI{10}{\kelvin}$. The data for Cu(2)/NiFe($d_{\mathrm{NiFe}}$)/Cu(2) samples were taken from Ref. \cite{Mishra2022}.}
\label{fig:MvH}
\centering
\end{figure}

The moment per unit area ($m/\mathrm{area}$) versus field ($H$) for a selected set of Pd(2)/NiFe($d_{\mathrm{NiFe}}$)/Pd(2) and Cu(2)/NiFe($d_{\mathrm{NiFe}}$)/Cu(2) sample are shown in Fig \ref{fig:MvH}. The curves look similar, but the $m/\mathrm{area}$ values in Pd samples are higher when compared to Cu samples with the same NiFe thickness. This is expected because Pd is partially magnetically polarized by the adjacent NiFe\cite{Smith2008}.

\begin{figure}[!htbp]
\includegraphics[width=\linewidth]{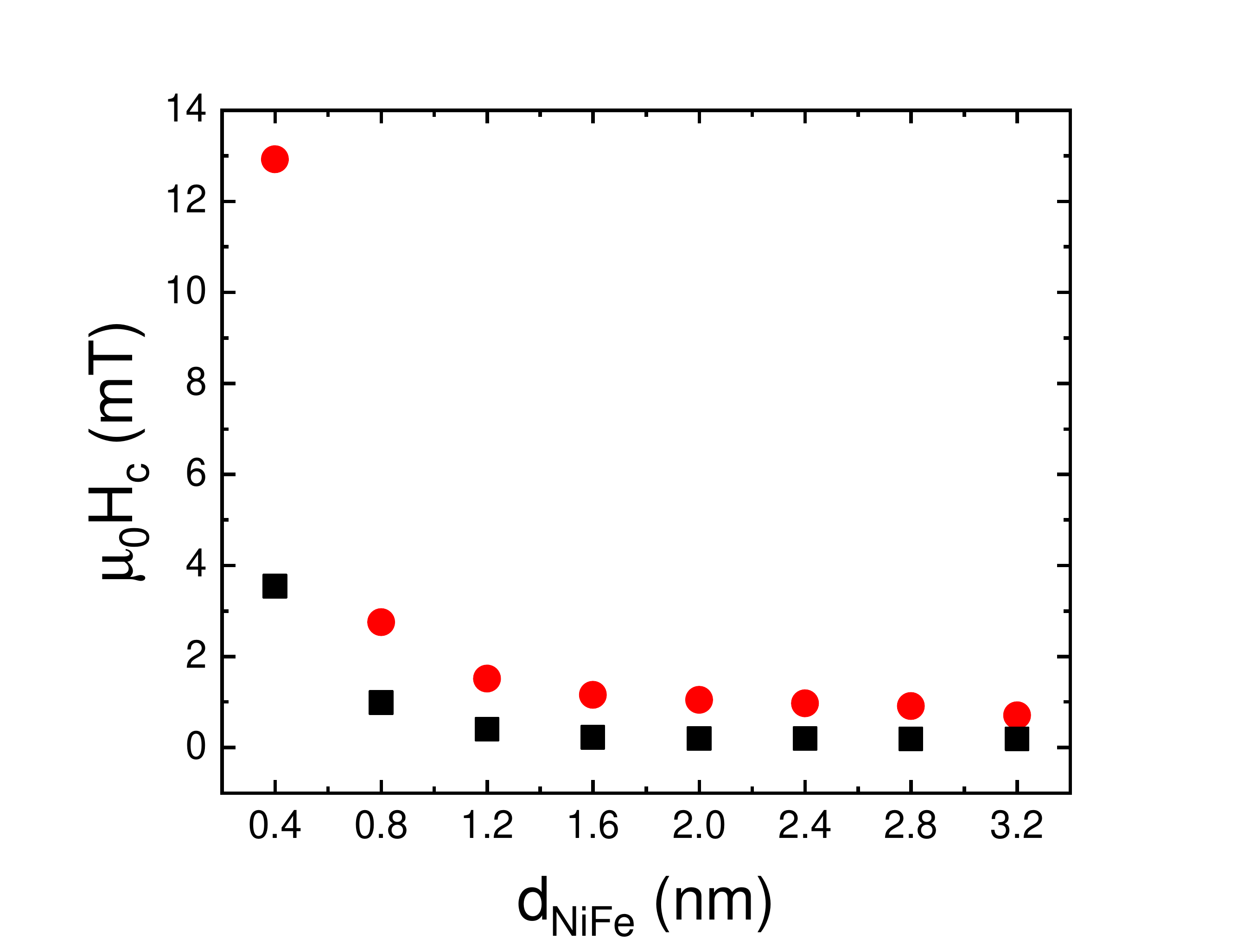}
\caption{Coercivity ($H_c$) vs NiFe thickness ($d_{\mathrm{NiFe}}$) for Cu(2)/NiFe($d_{\mathrm{NiFe}}$)/Cu(2) (black squares) and Pd(2)/NiFe($d_{\mathrm{NiFe}}$)/Pd(2) (red circles) films, measured at $T = \SI{10}{\kelvin}$.}
\label{fig:Hc_Comparison}
\centering
\end{figure}

The coercivities versus NiFe thickness for the Pd(2)/NiFe($d_{\mathrm{NiFe}}$)/Pd(2) and Cu(2)/NiFe($d_{\mathrm{NiFe}}$)/Cu(2) samples are plotted in Fig. \ref{fig:Hc_Comparison}. The coercivities for the Pd samples are somewhat higher than the Cu samples, however, the difference is not huge except for the extremely thin Pd(2)/NiFe(0.4)/Pd(2). Since magnetic switching in blanket films is aided by domain wall motion, coercivities for blanket films are generally lower than that of the small single-domain magnetic pillars in our Josephson junctions. Clearly the thinnest Pd(2)/NiFe(0.4)/Pd(2) trilayer could not be used as a free magnetic layer, but it might serve well as a fixed layer.
 
\begin{figure}[!htbp]
\includegraphics[width=\linewidth]{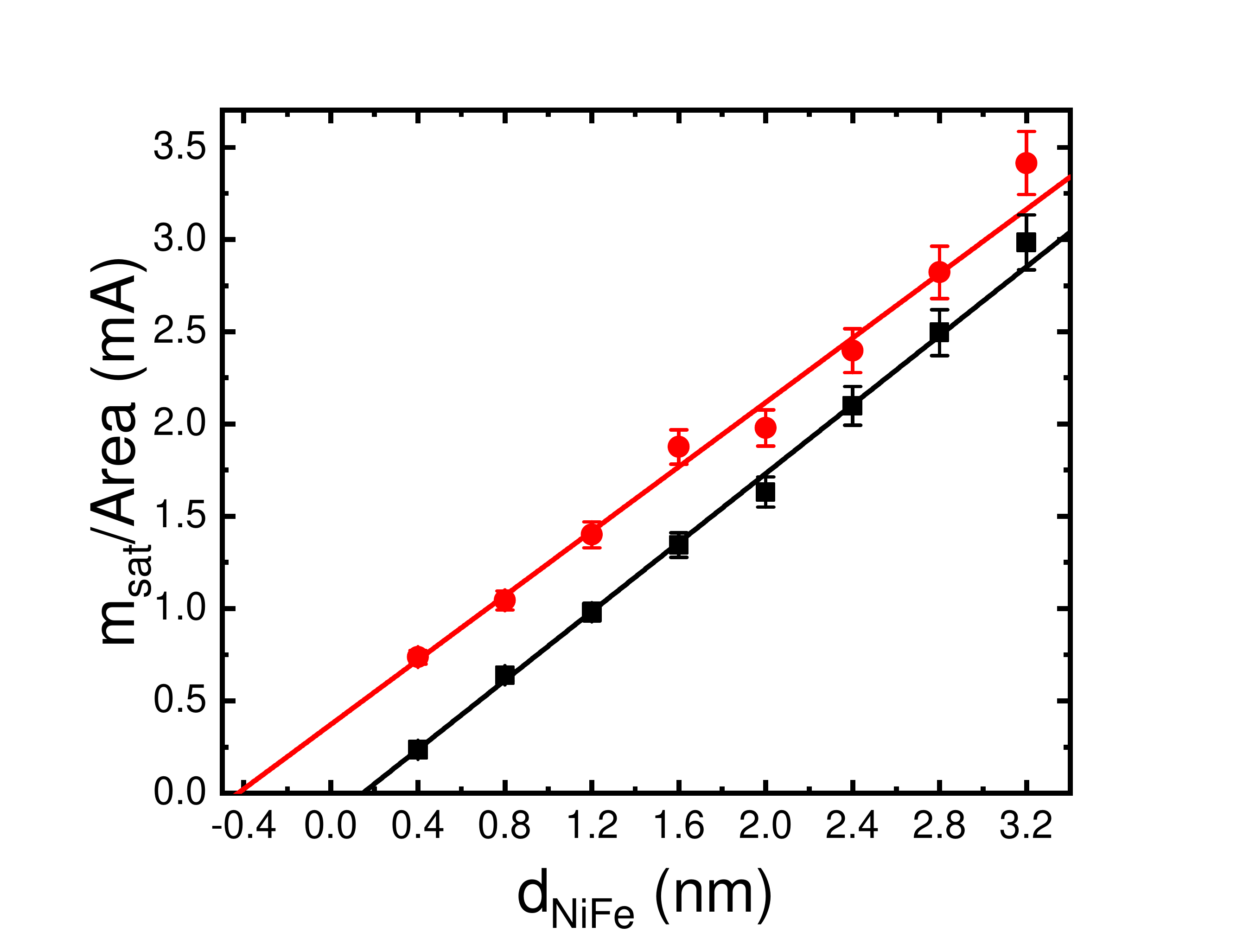}
\caption{Saturation moment per unit area vs NiFe thickness for Pd(2)/NiFe($d_{\mathrm{NiFe}}$)/Pd(2) (red circles) and Cu(2)/NiFe($d_{\mathrm{NiFe}}$)/Cu(2) (black squares) films, measured at $T = \SI{10}{\kelvin}$. The solid lines are linear fits.}
\label{fig:Msat}
\centering
\end{figure}

The saturation moment per unit area ($m_{sat}/\mathrm{Area}$) vs NiFe thickness ($d_{\mathrm{NiFe}}$) for Pd(2)/NiFe($d_{\mathrm{NiFe}}$)/Pd(2) and Cu(2)/NiFe($d_{\mathrm{NiFe}}$)/Cu(2) samples is shown in Fig. \ref{fig:Msat}. An uncertainty of 5\% is attributed to each data point, arising from the area estimation made using an optical microscope. Using straight line fits, the values of $M_{\mathrm{NiFe}}$ are determined from the slopes to be \SI{872}{} $\pm$ \SI{34}{kA/m} for the samples with Pd and \SI{934}{} $\pm$ \SI{16}{kA/m} for the samples with Cu. Those values are reasonably consistent with O’Handley’s low-temperature value of \SI{930}{kA/m} for Ni$_{80}$Fe$_{20}$ \cite{OHandley2000}, given the lower Fe concentration of our Permalloy.  The values of the x-intercepts from the fits suggest that the samples with Cu have a small dead layer thickness of about \SI{0.15}{\nm} (including both NiFe/Cu interfaces), while the Pd samples show a negative x-intercept of \SI{0.43}{\nm}, indicating that the Pd is partially polarized by promixity with NiFe.  Equivalently, the samples with Pd act as if each side of the NiFe layer has an extra thickness of \SI{0.21}{\nm}.

\subsection{Josephson junction transport}
Ferromagnetic Josephson junctions without an insulating barrier show overdamped dynamics. In such cases, the current-voltage ($I-V$) curves can be fit to the Resistively Shunted Junction model\cite{BaronePaterno1982}:
\begin{equation}
    V = \mathrm{sign}(I) R_N \Re\left\{\sqrt{I^2-I_c^2}\right\}
\end{equation}
where $R_N$ is the normal-state resistance of the junction, $\Re$ represents the real part of the argument and $I_c$ is the critical current. $I_c$ and $R_N$ can be obtained by fitting the equation above to the experimental data.

\begin{figure}[!htbp]
\includegraphics[width=\linewidth]{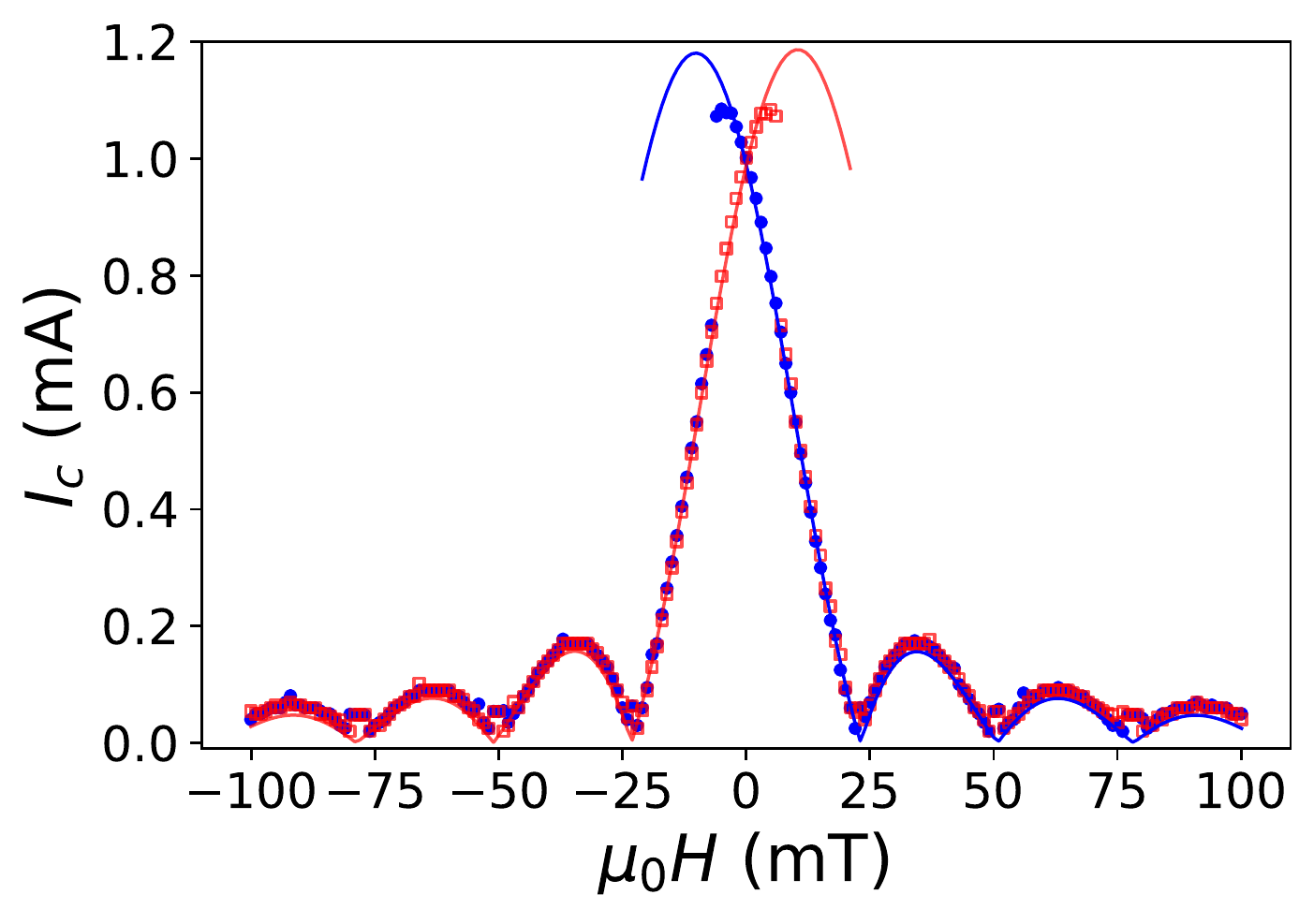}
\caption{Critical current vs field for Josephson junctions containing Pd(2)/NiFe(1.2)/Pd(2). The red squares and blue circles represent the data taken during magnetic field upsweep and downsweep, respectively. The solid lines are fits to Eqn. \ref{Eqn:Airy}. The sample is representative of all the samples in the set. However, the periods of the Airy patterns vary by $\pm$15-20\%, we believe due to variation in the junction widths during fabrication, which is discussed in the Appendix of Ref. \cite{Mishra2022}.}
\label{fig:Fraunhofers}
\centering
\end{figure}

For a Josephson junction, the critical current $I_c$ varies with a magnetic field $H$ applied in a direction perpendicular to the supercurrent flow. The dependence of $I_c$ on $H$ for a representative Josephson junction containing Pd(2)/NiFe(1.2)/Pd(2) is shown in Fig. \ref{fig:Fraunhofers}. The data in blue circles and red squares were acquired during the field downsweep and upsweep, respectively.

The experimental data are expected to follow an Airy function \cite{BaronePaterno1982} for elliptically shaped junctions when the field is applied along a principal axis:
\begin{equation} \label{Eqn:Airy}
    I_c(\Phi) = I_{c0} \left| \frac{2 J_1 \left( \frac{\pi \Phi}{\Phi_0} \right) }{\frac{\pi \Phi}{\Phi_0}} \right|
\end{equation}
where $J_1$ is the Bessel function of the first kind, $I_{c0}$ is the maximum value of $I_c$ through the junction and $\Phi_0 = \SI{2e-15}{\tesla\meter^2}$ is the flux quantum. The total magnetic flux through the junction is given by:
\begin{equation} \label{Eqn:Flux}
    \Phi = \mu_0 H w (2\lambda_{\mathrm{eff}} + d_N + d_F) + \mu_0 M w d_F
\end{equation}
where $w$ is the width of the junctions transverse to the field direction, $\lambda_{\mathrm{eff}}$ is the effective London penetration depth, $d_N$ is the thickness of the normal (non-ferromagnet/non-superconductor) layers, $d_F$ is the thickness of the ferromagnetic layers and $M$ is a weighted average of the NiFe and adjacent polarized Pd magnetizations. The solid lines in Fig. \ref{fig:Fraunhofers} are fits of Eqn. \ref{Eqn:Airy} to the experimental data for each field sweep. The center of the Airy pattern exhibits a hysteretic shift in either direction due to the internal magnetization of the ferromagnetic layers in the junction. The discontinuous drops in $I_c$ near $\pm$ \SI{10}{\milli\tesla} are due to switching of the magnetization direction. The field shift and switching fields vary with the NiFe thickness as they are due to inherent properties of the ferromagnetic layer. Because of the switch discussed above, the value of $I_{c0}$ extracted from the fits is typically higher than the maximum measured value. The ``true'' value is extracted from the fit, as shown by the extended fit lines in Fig. \ref{fig:Fraunhofers}.

\begin{figure*}[!htbp]
\includegraphics[width=\linewidth]{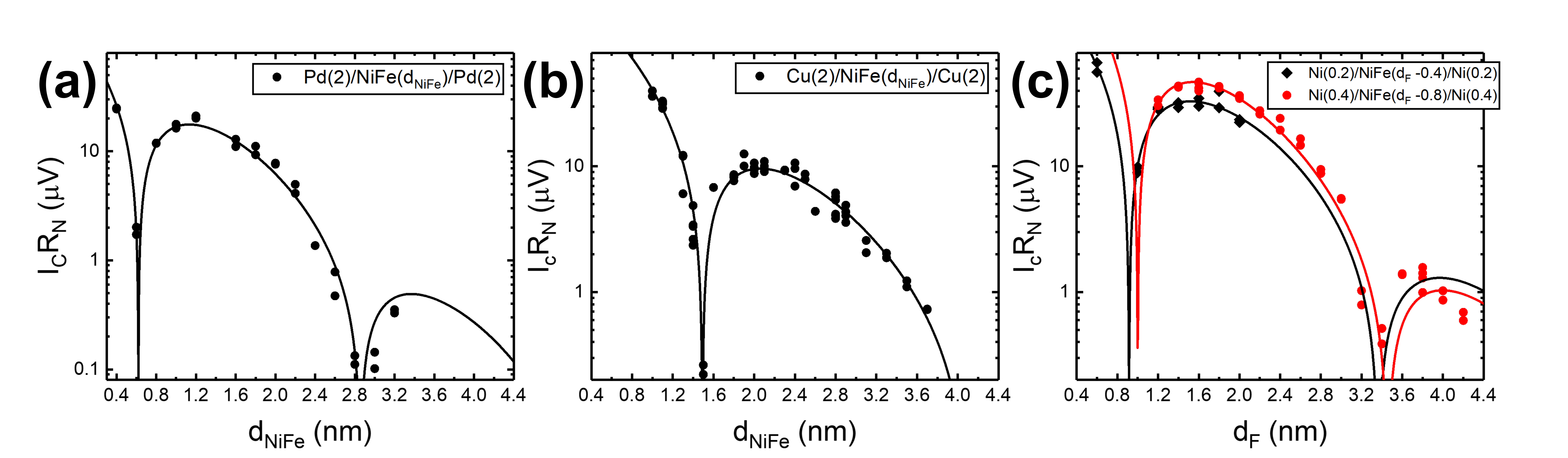}
\caption{$I_c R_N$ vs NiFe thickness $d_{\mathrm{NiFe}}$ for (a) Pd(2)/NiFe($d_{\mathrm{NiFe}}$)/Pd(2) and (b) Cu(2)/NiFe($d_{\mathrm{NiFe}}$)/Cu(2). (c) $I_c R_N$ vs total F-layer thickness $d_F = d_{\mathrm{Ni}} + d_{\mathrm{NiFe}}$ for junctions containing either Cu(2)/Ni(0.4)/NiFe($d_{\mathrm{NiFe}}$)/Ni(0.4)/Cu(2) or Cu(2)/Ni(0.2)/NiFe($d_{\mathrm{NiFe}}$)/Ni(0.2)/Cu(2). The solid lines in all panels are fits to Eqn. \ref{Eqn:Diffusive}. The data in (b) and (c) panels were published previously in Ref. \cite{Mishra2022}.}
\label{fig:ICRN}
\centering
\end{figure*}

Since $I_{c0}$ is proportional to the junction area, we multiply it by the normal state resistance $R_N$ to obtain the value of $I_c R_N$ for each sample. $I_c R_N$ is independent of junction area variations that can arise from fabrication inconsistencies. Figure \ref{fig:ICRN}(a) shows $I_c R_N$ versus NiFe thickness $d_{\mathrm{NiFe}}$ for all the Pd(2)/NiFe($d_{\mathrm{NiFe}}$)/Pd(2) Josephson junctions. Panel (b) of the figure shows comparison data from junctions containing Cu(2)/NiFe($d_{\mathrm{NiFe}}$)/Cu(2), and panel (c) shows data from our previous study of junctions containing either Cu(2)/Ni(0.4)/NiFe($d_{\mathrm{NiFe}}$)/Ni(0.4)/Cu(2) or Cu(2)/Ni(0.2)/NiFe($d_{\mathrm{NiFe}}$)/Ni(0.2)/Cu(2) \cite{Mishra2022}. All sample sets show a decaying oscillatory behavior of $I_c R_N$ with the thickness of the ferromagnet. The maximum value of $I_c R_N$ in the $\pi$-state of the Pd/NiFe/Pd junctions is about twice as high as that of the the Cu/NiFe/Cu junctions, while the maximum value for the Cu/Ni/NiFe/Ni/Cu junctions is about 4 times higher than that of Cu/NiFe/Cu junctions. Those are the main results of this work.

The dependence of $I_c R_N$ on the ferromagnet thickness has been calculated theoretically \cite{Buzdin2005} and measured experimentally by several groups for different ferromagnetic materials\cite{Ryazanov2001, Kontos2002, Blum2002, Sellier2003}. The behavior of $I_c R_N$ versus ferromagnet thickness is predicted to oscillate and decay, either algebraically for ballistic transport \cite{Buzdin1982} or exponentially for diffusive transport \cite{Buzdin1991}. The $I_c R_N$ data shown in Figure \ref{fig:ICRN}(a) oscillate and decay exponentially for all four data sets, hence we fit the data with the generic fitting function: 
\begin{equation} \label{Eqn:Diffusive}
    I_c R_N = V_0\; \mathrm{exp}\left(\frac{-d_F}{\xi_{F1}}\right) \left| \mathrm{sin} \left( \frac{d_F - d_{0-\pi}}{\xi_{F2}} \right)  \right|
\end{equation}
where $V_0$ is the magnitude of $I_c R_N$ extrapolated to zero F-layer thickness, $\xi_{F1}$ and $\xi_{F2}$ are length scales that control the decay and oscillation period in the ferromagnet F, and $d_{0-\pi}$ is the thickness where the first $0-\pi$ transition occurs. The solid lines in Fig. \ref{fig:ICRN} are fits of Eqn. \ref{Eqn:Diffusive} to the data with experimental uncertainties obtained from the Airy function fits. The uncertainties are smaller than the symbol size in Fig. \ref{fig:ICRN} and not visible. The fit parameters for all 3 data sets are tabulated in Table \ref{tab:parameters}.

The meaning of the length scales $\xi_{F1}$ and $\xi_{F2}$ depends on the relative sizes of the relevant energy scales (or equivalently, length or time scales) in the system.  The relevant parameter is $E_{\mathrm{ex}} \tau/\hbar$, where $2E_{\mathrm{ex}}$ is the energy splitting between the majority and minority spin bands in F and $\tau$ is the mean free time between collisions \cite{Bergeret2001}. In the case of weak ferromagnets in the diffusive limit, $E_{\mathrm{ex}} \tau/\hbar << 1$, the oscillatory-decay behavior can be calculated using the Usadel equation \cite{Buzdin1991,Buzdin2005}, with the result that both lengths should be equal to the dirty-limit exchange length, $\xi_F^* = \sqrt{\hbar D/E_{\mathrm{ex}}}$, where $D = v_F \tau/3$ is the diffusion constant in F.  Including spin-flip and/or spin-orbit scattering in the calculation causes $\xi_{F1}$ to shrink and $\xi_{F2}$ to lengthen \cite{Faure2006}.  However in the case of strong ferromagnets with $E_{\mathrm{ex}} \tau/\hbar > 1$, the Usadel equation is not strictly valid. A formula more appropriate for strong magnetic materials was derived by Bergeret \textit{et al.} using the Eilenberger equation \cite{Bergeret2001}; their formula also predicts an exponentially-decaying, oscillating function, but with the decay length $\xi_{F1}$ given by the mean free path and the oscillation length $\xi_{F2}$ equal to the clean limit exchange length $\xi_F = \hbar v_F/2E_{\mathrm{ex}}$.

To estimate which regime our samples are in, we use photoemission data obtained from NiFe by Petrovykh \textit{et al.} \cite{Petrovykh1998}. Those authors find $2E_{\mathrm{ex}} = \SI{0.27}{eV}$, $v_F = \SI{2.2e5}{\meter^2/\second}$, and a very short mean free path for minority electrons of $\SI{0.4}{}-\SI{0.8}{\nm}$. Choosing the mid-point of their mean free path range, $l=\SI{0.6}{\nm}$, we obtain $\xi_F=\SI{0.54}{\nm}$ and $\xi_F^*=\SI{0.46}{\nm}$, so all three length scales are nearly equal and close to the values of the fit parameters $\xi_{F1}$ and $\xi_{F2}$ given in Table \ref{tab:parameters}. We also find $E_{ex} \tau/\hbar=0.57$, right at the boundary between the two regimes of validity. 

We note that Robinson \textit{et al.} \cite{Robinson2006} found the somewhat longer value $\xi_{F1}=\SI{1.4}{\nm}$ for NiFe inside Josephson junctions, and interpreted it as the mean free path in their NiFe. It is plausible that the mean free path in sputtered NiFe varies by a factor of two in different laboratories, although more data from both our group and theirs would be needed to make a definitive comparison.

\section{Discussion}

The enhancement of the critical current of NiFe Josephson junctions by replacing the Cu/NiFe interfaces with Pd/NiFe appears to validate our original hypothesis based on spin-dependent scattering parameters derived from GMR data. We demonstrated similar results by replacing the Cu/NiFe interface with Cu/Ni/NiFe interfaces in a previous study\cite{Mishra2022}. For the latter, we suggested that the minority-band Fermi surface of Ni provides a better match to the Fermi surface of Cu than the NiFe minority-band does. We explained this phenomena by using the ``two-current series resistor'' model for calculating GMR parameters in the current-perpendicular-to-plane (CPP) geometry\cite{Zhang1991,Lee1993,Valet1993}. In this model, the transport through an F/N interface can be described by the following parameters: the dimensionless interface scattering asymmetry $\gamma_{F/N} = (AR^{\uparrow}_{F/N}-AR^{\downarrow}_{F/N})/(AR^{\uparrow}_{F/N}+AR^{\downarrow}_{F/N})$ and twice the enhanced interface specific resistance $2AR^*_{F/N}= (AR^{\uparrow}_{F/N} + AR^{\downarrow}_{F/N})/2$ where $AR_{F/N}^{\uparrow}$ and $AR_{F/N}^{\downarrow}$ are the interface specific resistances for the conduction electron moment pointing parallel and antiparallel to the F-layer magnetization, respectively. Large absolute values of $\gamma_{F/N}$ imply that one spin species has much poorer interface transmission than the other, which will be detrimental to the transmission of spin-singlet electron pairs. Similarly, a high value of $2AR^*_{F/N}$ indicates a low average interface transmission for electrons of either spin, again detrimental to pair transmission. Both $2AR^*_{F/N}$ and $\gamma_{F/N}$ for the Cu/Ni interface are lower compared to the Cu/NiFe interface. The same is also true for the Pd/NiFe interface compared to the Cu/NiFe interface. We tabulate both parameters in Table \ref{tab:interface}\cite{Bass2016}.

\begin{table*}[!htbp]
\caption{Parameters determined from fits of Eqn. \ref{Eqn:Diffusive} to the experimental data shown in Fig. \ref{fig:ICRN} for Josephson junctions containing Pd(2)/NiFe($d_{\mathrm{NiFe}}$)/Pd(2) , Cu(2)/NiFe($d_{\mathrm{NiFe}}$)/Cu(2), Cu(2)/Ni(0.4)/NiFe($d_{\mathrm{NiFe}}$)/Ni(0.4)/Cu(2) and Cu(2)/Ni(0.2)/NiFe($d_{\mathrm{NiFe}}$)/Ni(0.2)/Cu(2).}
\label{tab:parameters}
    \centering
    \begin{tabular}{|c|c|c|c|c|}
\hline
Sample & $V_0$ (\SI{}{\micro \volt}) & $\xi_{F1}$ (\SI{}{\nm}) & $\xi_{F2}$ (\SI{}{\nm}) & $d_{0-\pi}$ (\SI{}{\nm})\\
 \hline
Pd(2)/NiFe($d_{\mathrm{NiFe}}$)/Pd(2) & $164\pm6$ & $0.63\pm0.02$ & $0.71\pm0.01$ & $0.62\pm0.01$\\
Cu(2)/NiFe($d_{\mathrm{NiFe}}$)/Cu(2) & $252\pm48$ & $0.71\pm0.04$ & $0.74\pm0.06$ & $1.49\pm0.01$\\ 
Cu(2)/Ni(0.4)/NiFe($d_F$-0.8)/Ni(0.4)/Cu(2) &  $800\pm110$ & $0.64\pm0.02$ & $0.78\pm0.01$ & $1.00\pm0.02$\\
Cu(2)/Ni(0.2)/NiFe($d_F$-0.4)/Ni(0.2)/Cu(2) &  $349\pm39$ &  $0.76\pm0.05$ & $0.78$ (fixed) & $0.92\pm0.01$\\
 \hline
    \end{tabular}
\end{table*}

\begin{table}[!htbp]
\caption{Interface spin scattering asymmetry $\gamma_{F/N}$ and interface resistance $2AR^*_{F/N}$ for NiFe/Cu and Ni/Cu interfaces obtained from GMR studies\cite{Bass2016}.}
\label{tab:interface}
    \centering
    \begin{tabular}{|c|c|c|}
\hline
F/N Interface & $\gamma_{F/N}$ & $2AR^*_{F/N}$ (\SI{}{\femto\Omega\meter^2})\\
\hline
NiFe/Pd & 0.41 & 0.4\\
NiFe/Cu & 0.7 & 1.0\\
Ni/Cu & 0.3 & 0.36\\
\hline
    \end{tabular}
\end{table}

There is one feature in the data that we find puzzling: the Pd/NiFe/Pd data in Fig. \ref{fig:ICRN}(a) exhibit a $0-\pi$ transition at a NiFe thickness of about \SI{0.6}{\nm}, which when added to the effective polarized Pd thickness of \SI{0.42}{\nm}, gives the result \SI{1.0}{\nm}, which is similar to that of Ni\cite{Baek2018}, but much shorter than the values of \SI{1.5}{}-\SI{1.7}{\nm} observed in NiFe\cite{Glick2017,Dayton2018,Mishra2022}. (Note that the very small magnetic dead layer at the NiFe/Cu interface is much too small to explain the discrepancy.) We observed a similar unexplained effect in our Cu/Ni/NiFe/Ni/Cu data. The second transition shown in Fig. \ref{fig:ICRN}(a), which occurs at a NiFe thickness of about \SI{2.8}{\nm} (i.e. a total effective F-layer thickness at \SI{3.2}{\nm}) is similar to what is observed in NiFe junctions\cite{Baek2018,Glick2017,Dayton2018}. This seems to indicate that the interface plays a strong role in the location of the $0-\pi$ transition\cite{Heim2015, Pugach2011}. The switch from Cu to Pd also means that two Nb/Cu interfaces have been replaced by Nb/Pd interfaces. The properties of these Nb/Pd interfaces and their effect on the overall supercurrent transmission is unknown.

To solve all of these puzzles would require theoretical calculations of supercurrent that incorporate realistic modeling of the complex band structure of the strong ferromagnetic materials in the junctions.  Such calculations have been carried out only for the case of Nb/Ni/Nb junctions, by Ness \textit{et al.}\cite{Ness2022}. Unfortunately, such time-consuming calculations have not been carried out for other materials such as NiFe, or for the material combinations found in our junctions.

\section{Conclusion}
In this work, we show that the supercurrent transmission through Josephson junctions containing Cu/NiFe interfaces can be improved by replacing the Cu/NiFe interfaces with Pd/NiFe. The magnetic properties of Pd/NiFe/Pd are similar to Cu/NiFe/Cu with some differences: slightly higher coercivities and partial magnetic polarization of the Pd layers. We observe a factor of 2 increase in the supercurrent transmission of our NiFe Josephson junctions in the $\pi$-state when the adjacent Cu layers are replaced with Pd. We speculate that the reason behind this improvement could be better band-matching and lower resistance at the Pd/NiFe interface compared to the Cu/NiFe interface. However, to confirm the exact reason and quantify this improvement, theoretical band-structure calculations need to be performed.

\section*{Acknowledgments}
The authors thank V. Aguilar, T. F. Ambrose, R. M. Klaes, M. G. Loving, A. E. Madden, D. L. Miller, N. D. Rizzo, and J. C. Willard for their comments and insights. We also thank D. Edmunds and B. Bi for technical assistance with equipment. We acknowledge using the W. M. Keck Microfabrication Facility and other laboratories at Michigan State University for this study. This research was supported by Northrop Grumman Corporation.

\bibliography{IEEEabrv,references}

\end{document}